\documentclass[aps,prl,superscriptaddress,reprint,twocolumn]{revtex4-2}
\usepackage{tikz-cd}
\usetikzlibrary{graphs}
\usetikzlibrary{decorations.markings}
\usepackage{hyperref}
\hypersetup{
	breaklinks=true,   
	colorlinks=true,   
	pdfusetitle=true,  
}
\hypersetup{pdfborder={0 0 0},colorlinks=true,urlcolor=blue,citecolor=blue,linkcolor=blue,linktocpage=true}
\usepackage{amsmath,amssymb}
\usepackage{bm}
\usepackage{mathtools}
\newcommand{\beqv}[3]{\beta^{\rm eqv} \! \left[\begin{smallmatrix}#1\\#2\end{smallmatrix};#3\right]}
\newcommand{\beqvno}[2]{\beta^{\rm eqv}\! \left[\begin{smallmatrix}#1\\#2\end{smallmatrix}\right]}

\newcommand{\RRe}{\mathrm{Re}}
\newcommand{\zp}[1]{\frac{\zeta_{#1}'}{\zeta_{#1}}}
\newcommand{\zpt}[1]{\tfrac{\zeta_{#1}'}{\zeta_{#1}}}
\newcommand{\ZDG}[1]{Z^{\text{DG}}_{#1}}

\begin{document}
	\title{{Type II superstring amplitude at one-loop and transcendentality}}
	\author{Emiel Claasen}
	\affiliation{Max-Planck-Institut f\"ur Gravitationsphysik (Albert-Einstein-Institut), Am M\"uhlenberg 1, DE-14476 Potsdam, Germany}
	\author{Mehregan Doroudiani}
	\affiliation{Max-Planck-Institut f\"ur Gravitationsphysik (Albert-Einstein-Institut), Am M\"uhlenberg 1, DE-14476 Potsdam, Germany}
	\affiliation{School of Physics \& Astronomy, University of Southampton, SO17 1BJ, UK}
	\date{\today}
	\begin{abstract}
		{We calculate the four-graviton scattering amplitude in Type II superstring theory at one loop up to seventh order in the low-energy expansion through the recently developed iterated integral formalism of Modular Graph Functions (MGFs). The machinery of the novel method allows us to propose a general form of the amplitude, which suggests that the expansion is expressible in terms of single-valued multiple zeta values and logarithmic derivatives of the Riemann zeta function at positive and negative odd integers. Furthermore, we comment on the transcendental behavior of the amplitude.}
	\end{abstract}
	\maketitle
	\textit{Introduction.—} Low-energy expansions of superstring amplitudes reveal rich mathematical structures. For example, already at tree level, the expansion coefficients involve specific types of transcendental numbers. In particular, for open strings we find Multiple Zeta Values (MZVs) \cite{Schlotterer:2012ny}, defined by
	\begin{align}
		\zeta_{n_1, n_2, \ldots, n_r} \coloneqq \sum_{1 \leq k_1<k_2<\ldots<k_r}^{\infty} k_1^{-n_1} k_2^{-n_2} \dots k_r^{-n_r}\, ,
	\end{align}
	for $n_1,\ldots,n_r\in \mathbb{N}$ and $n_r>1$ ensuring convergence. For closed strings at tree level, only specific combinations of MZVs, called single-valued MZVs, appear \cite{Schlotterer:2012ny,Stieberger:2013wea,Stieberger:2014hba,Schlotterer:2018zce,Vanhove:2018elu,Brown:2019wna}. (Single-valued) MZVs correspond to specific values of iterated integrals of logarithmic forms and it is common in the physics literature to classify such iterated integrals according to the number of iterations. This quantity is also known as transcendental weight and is represented by the symbol $\mathcal{T}(f)$ for a large class of iterated integrals $f$, evaluated at algebraic values \cite{GONCHAROV201379,Brown:2009qja,Henn:2013pwa}. We have for example $\mathcal{T}(\mathrm{log})=1$ and $\mathcal{T}(\mathrm{Li}_n)=n$. We also assign transcendental weight to specific values, such as $\mathcal{T}(\pi)=1$ and $\mathcal{T}(\zeta_{n_1,n_2,\dots,n_r})=n_1+n_2+\dots+n_r$. Furthermore, from the definition of $\mathcal{T}$ it follows that $\mathcal{T}(fg)=\mathcal{T}(f)+\mathcal{T}(g)$ for $f,g$ (specific values of) iterated integrals. Assigning transcendental weight to evaluations of iterated integrals at transcendental values is not always well-defined, for example $\mathcal{T}(\log(\pi))$ is ambiguous.

	For tree-level amplitudes in flat spacetime of type I/II superstring theories, the transcendental weights of the expansion coefficients match (possibly up to an overall shift) with the expansion order — a property known as \textit{uniform transcendentality} \cite{Schlotterer:2012ny}. Moreover, the low-energy effective action of Type II superstring theory can be formulated in terms of the low-energy expansion of closed-string amplitudes. The zeroth order corresponds to the Einstein-Hilbert action and higher orders provide stringy corrections beyond general relativity that correspond to higher derivatives of the Riemann curvature tensor $D^{k}\mathcal{R}^n$. Consequently, the low-energy effective action of Type II superstring theory at tree level inherits uniform transcendentality.
	
	Advancing from tree-level to one-loop and specializing to closed strings, we find that the string worldsheet is conformally equivalent to a punctured torus with the punctures representing external states. Calculation of the one-loop amplitude involves an integral over the moduli space of the punctured torus, which can be split into integrals over the puncture positions followed by an integral over the moduli space of the torus. The configuration-space integral over the puncture positions, referred to in this letter as \textit{the integrand}, has a low-energy expansion in terms of a specific class of $\mathrm{SL}(2,\mathbb{Z})$-invariant functions of the torus modulus $\tau$. These are known as Modular Graph Functions (MGFs) and have been extensively studied in the literature \cite{Green:2000,Green:2008uj,DHoker:2015gmr,DHoker:2015sve,DHoker:2015wxz,DHoker:2016quv, Gerken:2018zcy,DHoker:2016mwo,Basu:2016kli,Kleinschmidt:2017ege}. Integration of MGFs over the moduli space of the torus introduces numbers such as the Euler-Mascheroni constant and the logarithmic derivative of the Riemann zeta function $\zeta_{z}$ \cite{DHoker:2019mib,DHoker:2019blr,DHoker:2021ous} ({the argument is promoted} to $z\in \mathbb{C}$), whose transcendentality properties remain poorly understood.
	
	{In this letter, employing the recently developed iterated integral formalism of MGFs \cite{Broedel:2018izr,Gerken:2018jrq,Gerken:2019cxz,Gerken:2020yii,Brown:2017qwo,Brown:2017qwo2,Dorigoni:2022npe,Dorigoni:2024oft,Doroudiani:2023bfw,Claasen:2025vcd}, we reproduce the amplitude calculation of \cite{DHoker:2019blr} up to sixth order and extend it to seventh order in the low-energy expansion, which corresponds to the $D^{14}\mathcal{R}^4$ term in the effective action. The new results disprove the conjectural general form of the low-energy expansion of the amplitude given in \cite{DHoker:2019blr}. {Moreover}, the novel method allows us to pose an improved general form. {The new results} also suggests a different transcendentality assignment compared to \cite{DHoker:2019blr}, which is contingent on a number of assumptions.}
	
	\textit{Four-graviton string scattering.—}Consider the four-graviton amplitude in Type II superstring theory in flat spacetime. Let $k_i$ denote the asymptotic momenta and $\epsilon_i$ the polarization tensors of the gravitons, with $i=1,2,3,4$. Momentum conservation and the masslessness condition imply $\sum_{i=1}^4 k_i = 0$ and $k_i^2 = 0$ respectively, leaving two independent momenta. We use the dimensionless kinematic variables defined by $s_{ij} = -\alpha^\prime(k_i + k_j)^2 / 4$, called the Mandelstam invariants. We set $s = s_{12}=s_{34}$, $t = s_{14}=s_{23}$, and $u = s_{13}=s_{24}$, subject to $s + t + u = 0$. The perturbative four-graviton amplitude is then expressed as \cite{Green:2008uj,DHoker:2015gmr}
	\begin{equation}
		{A}(k_i, \epsilon_i) = \kappa_{10}^2 \mathcal{R}^4 \sum_{h=0}^\infty g_s^{2h-2} {A}^{(h)}(s_{ij}) \, ,
		\label{eq: fullamplitude}
	\end{equation}
	where $\kappa_{10}$ is the ten-dimensional gravitational constant. The factor $\mathcal{R}^4$ represents a contraction of four linearized Riemann tensors constructed from $k_i$ and $\epsilon_i$, capturing all graviton polarizations. Further, $g_s$ is the string coupling constant, $h$ denotes the genus of the worldsheet, and ${A}^{(h)}(s_{ij})$ are the genus-$h$ coefficient functions that depend on the kinematic variables $s_{ij}$. These functions are invariant under permutations of the $s_{ij}$ variables, implying we can express their momentum dependence through symmetric polynomials $\sigma_k = s^k + t^k + u^k$ for $k \in \mathbb{N}$. Momentum conservation ensures $\sigma_1 = 0$, leaving only $\sigma_2$ and $\sigma_3$ as independent variables \cite{Green:2000}.
	
	At tree level, the coefficient function is given by \cite{Green:2000}
	\begin{equation}
		{A}^{(0)}(s_{ij})=\frac{3}{\sigma_3}\exp\left\{2\sum_{m=1}^\infty \frac{\zeta_{2m+1}}{2m+1}\sigma_{2m+1}\right\} \, ,
		\label{eq: tree-sigmas}
	\end{equation}
	while closed formulae of coefficient functions at higher genus remain elusive.~However, by expanding the integrand for small values of the kinematic variables, the coefficient functions expand accordingly and the contributions of the individual terms can be calculated through explicit integration. This procedure is known as the low-energy expansion of the string amplitude. Performing a low-energy expansion of (\ref{eq: tree-sigmas}) {and using $\mathcal{T}(\zeta_{2m+1})=2m+1$}, we observe that the expression is uniformly transcendental.
	
	At genus one, the external asymptotic string states correspond to punctures on the torus worldsheet. In order to determine the genus-one four-point coefficient function, we need to perform an integral over the moduli space of the four-punctured torus. The torus can be represented by the quotient $\Sigma_\tau = \mathbb{C}/\Lambda$, where $\Lambda = \mathbb{Z} + \tau \mathbb{Z}$ is a lattice defined by the complex structure modulus $\tau=\tau_1+i\tau_2$. The moduli space of the torus is $\mathcal{M}=\mathbb{H}/\text{PSL}(2,\mathbb{Z})$, where $\mathbb{H}$ denotes the complex upper half-plane, and $\text{PSL}(2,\mathbb{Z})$ is the modular group. To represent the moduli space, we choose a fundamental domain
	\begin{align}
		\mathcal{M}&=\{\tau\in\mathbb{H}\mid-\tfrac{1}{2}\leq \RRe(\tau)\leq 0,|\tau|^2\geq 1\}\nonumber\\&\qquad\cup\, \{\tau\in\mathbb{H}\mid 0\leq \RRe(\tau)<\tfrac{1}{2},|\tau|^2>1\} \, .
		\label{eq: funddom}
	\end{align}
	For later convenience, we partition the moduli space $\mathcal{M}$ into two complementary regions, defined as follows \cite{Green:2000}
	\begin{gather}
		\mathcal{M}=\mathcal{M}_L\cup\mathcal{M}_R \, ,\nonumber\\ \mathcal{M}_L=\mathcal{M}\cap \{\tau_2\leq L\}\,,\quad \mathcal{M}_R=\mathcal{M}\cap \{\tau_2>L\}\, ,
		\label{eq: funddomsplit}
	\end{gather}
	with an auxiliary cutoff $L>1$ that drops out from the final results.
	
	A key element in genus-one amplitude calculations is the Green function $G(z, \tau)$ on the torus. Being doubly-periodic, this function admits a double Fourier expansion 
	\begin{equation}
		G(z,\tau)=\frac{\tau_2}{\pi}\sum_p^\prime \frac{e^{2\pi i\langle p,z \rangle}}{|p|^2} \, ,
		\label{eq: greensfourier}
	\end{equation}
	with $p=m\tau+n$ for $m,n\in\mathbb{Z}$. $z$ is the coordinate of a puncture on the torus and $\langle p,z\rangle=\tfrac{p\Bar{z}-\Bar{p}z}{2i\tau_2}$. The restriction on the sum, denoted by a prime, means $(m,n)\neq (0,0)$.
	
	The splitting of the moduli space integrations results in the amplitude \cite{Green:1981yb}
	\begin{equation}
		{A}^{(1)}(s_{ij})=2\pi\int_{\mathcal{M}}\frac{d^2\tau}{\tau_2^2}\mathcal{B}(s_{ij}|\tau)\, ,
		\label{eq: g1amplitude}
	\end{equation}
	where the integrand is given by the integral over the configuration-space of the punctures (with one fixed by translation invariance)
	\begin{equation}
		\mathcal{B}(s_{ij}|\tau)=\prod_{k=2}^4\int_{\Sigma_\tau}\frac{d^2z_k}{\tau_2}\exp\left\{{\sum_{1\leq i<j\leq 4} s_{ij}G_{ij}(\tau)}\right\},
		\label{eq: integrandB}
	\end{equation}
	where we used shorthand notation $G_{ij}(\tau)=G(z_i-z_j|\tau)$. The exponential is often called the Koba-Nielsen factor. The amplitude (\ref{eq: g1amplitude}) decomposes under the partitioning of $\mathcal{M}$ as
	\begin{equation}
		{A}^{(1)}(s_{ij})=\mathcal{A}_L(L;s_{ij})+\mathcal{A}_R(L;s_{ij})\, ,
	\end{equation}
	where the subscripts of $\mathcal{A}_L$ and $\mathcal{A}_R$ imply that we only integrate $\mathcal{B}(s_{ij}|\tau)$ over $\mathcal{M}_L$ and $\mathcal{M}_R$ respectively. Furthermore, we expect all the $L$-dependence to cancel as the partitioning is arbitrary.
	
	In the low-energy expansion, the amplitude naturally separates into analytic and non-analytic terms in the kinematic variables. Consequently, a second splitting of the amplitude can be introduced \cite{Green:2000,Green:2008uj}
	\begin{equation}
		{A}^{(1)}(s_{ij})=\mathcal{A}_{\text{an}}(s_{ij})+\mathcal{A}_{\text{non-an}}(s_{ij})\, .
	\end{equation}
	This distinction will be useful when we present the results and analyze the transcendental behavior of the amplitude.~The non-analytic terms in the four-graviton one-loop amplitude originate from the integral over $\mathcal{M}_R$, which has been computed for all orders in \cite{DHoker:2019blr}. Therefore, we are only concerned with the expansion of the integral over $\mathcal{M}_L$.
	
	\textit{Modular graph functions.---}The low-energy expansion of (\ref{eq: integrandB}) manifests itself as a Taylor expansion of the exponential in the kinematic variables near zero. This yields
	\begin{equation}
		\mathcal{B}(s_{ij}|\tau)=\sum_{\ell=0}^\infty \sum_{\sum\ell_{ij}=\ell}\prod_{1\leq i<j}^4\frac{s_{ij}^{\ell_{ij}}}{\ell_{ij}!}D_{\underline{\ell}}(\tau)\, ,
		\label{eq: BMGF}
	\end{equation}
	where we defined $\underline{\ell}=(\ell_{12},\ell_{13},\ell_{14},\ell_{23},\ell_{24},\ell_{34})\in \mathbb{N}_0^6$ and
	\begin{equation}
		D_{\underline{\ell}}(\tau)=\prod_{k=2}^4\int_{\Sigma_\tau}\frac{d^2z_k}{\tau_2}\prod_{1\leq i<j}^4 G_{ij}^{\ell_{ij}}(\tau)\, .
		\label{eq: MGF}
	\end{equation}
	As (\ref{eq: BMGF}) is invariant under permutations of $s_{ij}$, we can rewrite it using the symmetric polynomials as
	\begin{equation}
		\mathcal{B}(s_{ij}|\tau)=\sum_{p,q=0}^\infty \mathcal{B}_{(p,q)}(\tau)\frac{\sigma_2^p\sigma_3^q}{p!q!}\, ,
		\label{eq: Bsigma}
	\end{equation}
	where the coefficients $\mathcal{B}_{(p,q)}(\tau)$ are $\mathbb{Q}$-linear combinations of different $D_{\underline{\ell}}(\tau)$. The integrals in (\ref{eq: MGF}) can be evaluated using the Fourier expansion (\ref{eq: greensfourier}), leading to the lattice-sum representation
	\begin{align}
		D_{\underline{\ell}}(\tau)&= \left(\frac{\tau_2}{\pi}\right)^{\sum_{i<j} \ell_{i j}} \sum_{\left\{p_{i j}^{(k)}\right\}}^{\prime} \prod_{i<j} \prod_{k=1}^{\ell_{i j}} \frac{1}{\left|p_{i j}^{(k)}\right|^2}\nonumber \\
		&\quad\times \delta\left(p_{12}+p_{23}+p_{24}\right) \delta\left(p_{13}+p_{23}+p_{34}\right) \nonumber \\ &\quad\times\delta\left(p_{14}+p_{24}+p_{34}\right) \, ,
		\label{eq: latticesum}
	\end{align}
	where $p^{(k)}_{ij}=m^{(k)}_{ij}\tau+n^{(k)}_{ij}$ with $m^{(k)}_{ij},n^{(k)}_{ij}\in\mathbb{Z}$, $p_{ij}=\sum_{k=1}^{\ell_{ij}} p^{(k)}_{ij}$, and $\delta$ is the Kronecker delta.
	
	There exists a convenient graphical representation of the $D_{\underline{\ell}}(\tau)$ from which the lattice-sum representation (\ref{eq: latticesum}) naturally follows \cite{Green:2008uj,DHoker:2015gmr}
	\begin{equation}
		D_{\underline{\ell}}=\begin{tikzpicture}[baseline={([yshift=-0.4ex]current bounding box.center)}]
			\draw (0,0) to node[midway, fill=white]{$\ell_{14}$} (3,0);
			\draw (0,0) to node[midway, fill=white]{$\ell_{12}$} (0,3);
			\draw (3,0) to node[midway, fill=white]{$\ell_{34}$} (3,3);
			\draw (0,3) to node[midway, fill=white]{$\ell_{23}$} (3,3);
			\draw (0,0) to node[pos=0.25, fill=white]{$\ell_{13}$} (3,3);
			\draw (0,3) to node[pos=0.75, fill=white]{$\ell_{24}$} (3,0);
			\fill (0,0) circle (2pt);
			\fill (3,0) circle (2pt);
			\fill (3,3) circle (2pt);
			\fill (0,3) circle (2pt);
			\node[below] at (0,0) {$1$};
			\node[below] at (3,0) {$4$};
			\node[above] at (0,3) {$2$};
			\node[above] at (3,3) {$3$};
		\end{tikzpicture}\,.
		\label{Fig:ModGraph}
	\end{equation}
	Here, each edge represents a factor $G_{ij}(\tau)$ in (\ref{eq: MGF}) and $\ell_{ij}$ is the number of edges between vertices $i$ and $j$. By interpreting the Green functions as propagators with discrete momenta $p^{(k)}_{ij}$ and imposing momentum conservation at each vertex, we are led to (\ref{eq: latticesum}). Furthermore, we define the weight of an MGF to be equal to the total number of edges, which coincides with the expansion order. From (\ref{eq: latticesum}) we see that the $D_{\underline{\ell}}(\tau)$ are modular functions. In particular, they form a class of modular functions called Modular Graph Functions (MGFs) owing to their graphical representations \cite{Green:2008uj,DHoker:2015gmr}.
	
	It is straightforward to calculate $\mathcal{B}(s_{ij}|\tau)$ at a certain order and find the linear combination of MGFs contributing to the amplitude at that order. As a result of the wealth of identities of MGFs \cite{DHoker:2015gmr,DHoker:2015sve,DHoker:2015wxz,DHoker:2016mwo,DHoker:2016quv, Gerken:2018zcy,Basu:2016kli,Kleinschmidt:2017ege}, only a handful of them remain at low orders in the expansion.
	
	\textit{Integration of Iterated Integrals.—}
	To calculate the contributions to the string amplitude, the remaining task is to integrate the MGFs over $\mathcal{M}_L$. However, the lattice-sum representations are not well-suited for such an integration. Therefore, we consider an alternative representation of MGFs through modular iterated integrals of holomorphic Eisenstein series. We denote these iterated integrals by $\beqv{j_1&\dots&j_\ell}{k_1&\dots &k_\ell}{\tau}$, where $\ell$ is the number of iterations (also known as depth), and the $j_i,k_i$ specify the integration kernels detailed in \cite{Dorigoni:2022npe,Dorigoni:2024oft}. We define $\beqv{}{}{\tau}=1$. The conversion from lattice sums to iterated integrals can be performed algorithmically \cite{Claasen:2025vcd}, which is inspired by the sieve algorithm \cite{DHoker:2016mwo,Gerken:2020aju}. Using this algorithm, we can expand MGFs in terms of $\mathbb{Q}[\text{single-valued MZV}]$-linear combinations of modular iterated integrals \cite{Zerbini:2015rss,DHoker:2015wxz,Dorigoni:2024oft}. An interesting observation is that setting single-valued MZVs to zero yields $\mathbb{Z}$-linear combination of $\beta^{\mathrm{eqv}}$'s in all examples.
	
	As an example, consider $D_5 \coloneq D_{(5,0,0,0,0,0)}(\tau)$, which can be written as
	\begin{align}
		&D_{5}=1800\big(-4\beqvno{3\,0}{6\,4}+\beqvno{1\,2}{4\,6}-4\beqvno{2\,1}{4\,6}\notag\\&+\beqvno{2\,1}{6\,4}-9\beqvno{4}{10}\big)-60\beqvno{1}{4}\zeta_3+15\zeta_5\, ,
	\end{align}
	where we omit the explicit $\tau$ dependence. We see that $D_5$ can be written in terms of depth 0, 1 and 2 modular iterated integrals. We provide explicit expansions of all MGFs appearing in the low-energy expansion up to seventh order in the ancillary file.
	
	One can construct a different basis of iterated integrals consisting of modular functions $F(\tau)$ satisfying inhomogeneous Laplace equations \cite{DHoker:2015sve,Dorigoni:2021jfr,Dorigoni:2021ngn,Dorigoni:2024oft}
	\begin{align}
		(\Delta-\mu)F(\tau) = \text{source} \, .
	\end{align}
	Here $\Delta=4\tau_2^2\partial_\tau\partial_{\Bar{\tau}}$ is the Laplace-Beltrami operator, $\mu\in\mathbb{Z}$, and the source terms are sums of products of $\beta^{\mathrm{eqv}}$'s. Integrals over $\mathcal{M}_L$ of this basis corresponding to various source terms were obtained up to depth 3 \cite{Doroudiani:2023bfw} using Stokes theorem and the Rankin-Selberg-Zagier method \cite{zagier1981rankin}. As MGFs of weight $w$ expand in terms of this basis up to depth $\lfloor w/2 \rfloor$, we can express all MGFs up to weight 7 using at most depth 3 iterated integrals. Although the integrals in \cite{Doroudiani:2023bfw} only cover functions for which $\mu \neq 0$, they still suffice to determine the four-graviton amplitude at genus one up to seventh order. The change of iterated integral basis can be found in the ancillary file of \cite{Dorigoni:2024oft}, while their integrals relevant to the amplitude can be found in the ancillary file of this letter.
	
	Adding the integral of MGFs over $\mathcal{M}_L$ to the corresponding integral over $\mathcal{M}_R$ from \cite{DHoker:2019blr}, we obtain the low-energy expansion of the amplitude up to seventh order.
	
	\textit{Results.—}As expected, the method using iterated integrals reproduces the results of \cite{DHoker:2019blr} and extends them to seventh order. The primary challenge of using iterated integrals lies in determining the constant term in the $\beta^{\mathrm{eqv}}$ expansion of MGFs, see for example \cite{Gerken:2020yii,Dorigoni:2024oft}. This constant term contributes to the analytic part of the amplitude and is expressed as a $\mathbb{Q}$-linear combination of single-valued MZVs.
	
	In particular, to calculate this term we need the constant terms in the $\tau_2 \to i\infty$ limit of MGFs. These are known up to weight 6 \cite{DHoker:2015gmr,DHoker:2016quv,Gerken:2020aju}, but their computation becomes unmanageable at higher weights such that, at weight 7, only a subset of them is known \cite{Zerbini:2015rss}. However, a correspondence with UV-divergences in effective field theory one-loop matrix elements provides a useful workaround \footnote[1]{We thank Oliver Schlotterer for bringing this method to our attention.}. The overall constant term for the combination of MGFs in the integrand can be inferred from the ${\alpha^\prime}^7\zeta_7$ divergence identified in Eq.~(6.33) of \cite{Edison:2021ebi}. This suffices to determine the combination of constants in the $\beta^{\mathrm{eqv}}$ expansion at weight 7. Integrating the constants over $\mathcal{M}_L$ gives
	\begin{align}
		\mathcal{A}_{\text{an}}(s_{ij})&=\frac{2\pi^2}{3}\bigg(1+\frac{\zeta_{3}\sigma_3}{3}+\frac{29\zeta_5\sigma_2\sigma_3}{180}+\frac{\zeta_{3}^2\sigma_3^2}{18}\nonumber\\&\quad\quad\quad\quad-\frac{163\zeta_7\sigma_2^2\sigma_3}{30240}+\mathcal{O}(s_{ij}^8)\bigg)\, .
		\label{eq: analytic}
	\end{align}
	The remaining integrals are contained in
	\begin{align}
		\mathcal{A}_{\text{non-an}}(s_{ij})&=\frac{2\pi^2}{3}\bigg(\hat{\mathcal{A}}_{\text{sugra}}+\hat{\mathcal{A}}_4+\hat{\mathcal{A}}_6\nonumber \\
		&\quad+\hat{\mathcal{A}}_7+\mathcal{O}(s_{ij}^8)\bigg)\, .
		\label{eq: nonanalytic}
	\end{align}
	The first lines of (\ref{eq: analytic}) and (\ref{eq: nonanalytic}) were already calculated in \cite{DHoker:2019blr}, which we reproduce using the new method. Note that there is a slight abuse of notation here, as the so-called “non-analytic part” also includes analytic contributions. MGFs do not contribute to the first term of (\ref{eq: nonanalytic}), which corresponds to the one-loop amplitude of 10-dimensional supergravity calculated in \cite{Green:1982sw,Green:2008uj}. The remaining terms involve $\zeta_z$ and its logarithmic derivative, the Euler-Mascheroni constant $\gamma_E$, and $\log \pi$. However, by rewriting the result using only odd $\zeta_z$ and its logarithmic derivatives through analytic continuation \footnote[2]{We thank Herbert Gangl and Vincent Maillot for suggesting this reflection.}
	\begin{align}
		\zp{2z}+\zp{1-2z} = \log (2\pi) + \gamma_E - H(2z-1)\, ,
		\label{reflectionEven}
	\end{align}
	most of these numbers cancel out. Here $H(z)$ denotes the harmonic sum $H(z) = \sum_{k=1}^z \frac{1}{k}$.
	The results are as follows
	\begin{align}
		\hat{\mathcal{A}}_4&=-\frac{4}{15} \zeta_{3} s^4 \log (-s)+\text{2 c.p. of $(s,t,u)$}\nonumber\\&\quad + \frac{2\zeta_3 \sigma_2^2}{15}\left[-\frac{\zeta_{-3}'}{\zeta_{-3}}-\frac{\zeta_{3}'}{\zeta_{3}}+\frac{79}{60}\right]\, ,
	\end{align}
	where “c.p.” stands for cyclic permutations.
	\begin{align}
		&\hat{\mathcal{A}}_6=-\tfrac{42s^6+s^4\sigma_2}{210}\zeta_5\log(- s)+2\,\text{c.p. of $(s,t,u)$}\nonumber\\&+\tfrac{\zeta_5\sigma_2^3}{630}\left[35\zpt{-1}-70\zpt{-3}+2\zpt{-5}-33\zpt{5}+\tfrac{13487}{420}\right]\nonumber\\&+ \tfrac{\zeta_5\sigma_3^2}{270}\left[-5\zpt{-1}+10\zpt{-3}-23\zpt{-5}-18\zpt{5}+\tfrac{3931}{210}\right] \, .
	\end{align}
	For weight 7 we find the new result
	\begin{align}                      &\hat{\mathcal{A}}_7=\tfrac{-26  s^7+s^5 \sigma _2}{210} \zeta_{3}^2 \log (-s)+ 2\,\text{c.p. of $(s,t,u)$}\\&+\frac{\zeta_{3}^2\sigma_2^2\sigma_3}{630}\left[-28\zpt{-3}-15\zpt{-5}-58\zpt{3}+15\gamma_E+\tfrac{50417}{1260}\right]\, .\nonumber
	\end{align}
	The logarithmic terms arise from the integration over $\mathcal{M}_R$ given in \cite{DHoker:2019blr}.
	
	\textit{General structure of the amplitude.—}The general form of the integral of a modular function over $\mathcal{M}_L$ was derived in \cite{Doroudiani:2023bfw} (see Eq.~(4.1) therein). Combining this result with the findings of \cite{DHoker:2019blr} for the integral over $\mathcal{M}_R$, it follows that the analytic part of the amplitude at each order is proportional to the constant term of the zero-modes of MGFs and $\beta^{\mathrm{eqv}}$’s. In both cases, this constant term is a single-valued MZV that preserves uniform transcendentality of the amplitude \cite{DHoker:2015wxz,Zerbini:2015rss,Dorigoni:2024oft}. Consequently, starting from weight 11, we expect the appearance of irreducible single-valued MZVs. The analytic part of the amplitude can be expressed schematically for all orders as
	\begin{align}
		\mathcal{A}_{\text{an}}(s_{ij}) = \frac{2\pi^2}{3}\sum_w \text{MZV}^{\text{sv}}_w \sigma_w \, ,
		\label{eq: AnGen}
	\end{align}
	where $\text{MZV}^{\mathrm{sv}}_w$ belongs to the $\mathbb{Q}$-algebra of single-valued MZVs with total transcendental weight $w$. 
	
	The non-analytic part is more intricate. The cancellation of $\log(L)$ upon combining the integrals over $\mathcal{M}_L$ and $\mathcal{M}_R$, along with the fact that the logarithmic dependence on Mandelstam variables always takes the form $\log(Ls)$, implies that the coefficient of $\log(-s)$ corresponds to the linear term in the zero-mode of $\beta^{\mathrm{eqv}}$’s in the expansion of MGFs. At order $w$, as proposed in \cite{Dorigoni:2024oft}, this coefficient is a single-valued MZV of transcendental weight $w-1$. By unitarity, being connected to the tree-level four-point amplitude, it is free of irreducible MZVs beyond depth~1.
	
	Up to weight 7, the non-analytic terms involve $\gamma_E$, $\zeta_z$ and its logarithmic derivatives, and rational numbers. Additionally, as noted in \cite{DHoker:2021ous, Doroudiani:2023bfw}, starting from weight 8, $\log(2\pi)$ appears in the integrals of MGFs. However, by using an appropriate reflection (\ref{reflectionEven}), this term can be eliminated. Using Eq.~(4.1) of \cite{Doroudiani:2023bfw}, the origin of the \(\zp{z}\) terms can be traced to the residue of the regularized Rankin-Selberg transform of MGFs (see Eqs. (2.17) and (2.18) of \cite{Doroudiani:2023bfw} for the definition) at 1. The regularized Rankin-Selberg transform of MGFs has been shown to be an L-function \cite{Zerbini:unpublished}. A consequence of Beilinson’s conjecture \cite{Beilinson} is that the residue of an L-function with a simple pole is a period. However, in the examples of these L-functions where the expansion around 1 is known, \(\zp{z}\) appears only when higher poles are present, leaving the precise nature of these residues unresolved. Notably, \(\zp{0} = \log(2\pi)\) \cite{Lerch}, which is the logarithm of a period.
	
	Overall, the schematic form of the non-analytic term for all orders is given by
	\begin{align}
		&\mathcal{A}_{\text{non-an}}= \frac{2\pi^2}{3}\bigg(\sum_w\left(\mathrm{RS}(w) + R_w \text{MZV}^{\mathrm{sv}}_{w-1}\right) \sigma_w \notag\\&+\hat{\mathcal{A}}_{\text{sugra}} + \sum_{w,k} \mathrm{ZV}^{\mathrm{sv}}_{w-1} s^{k}\sigma_{w-k}\log(-s) +2\,\text{c.p.}\bigg)\, ,
	\end{align}
	where $\mathrm{ZV}^{\mathrm{sv}}_w$ belongs to the $\mathbb{Q}$-algebra of single-valued Riemann zeta values with total transcendental weight $w$, $R_w\in\mathbb{Q}$, and $\text{RS}(w)$ represents contributions arising from the residue of the regularized Rankin-Selberg L-function.
	
	Up to weight 7, $\text{RS}(w)$ takes the following form
	\begin{align}
		\text{RS}(w) = \mathrm{ZV}^{\mathrm{sv}}_{w-1}\left( \sum_{n\in N} \alpha_n\zp{n} + \beta_w \gamma_E + \Tilde{R}_w\right)\, ,
		\label{eq: RankinSelberg}
	\end{align}
	where $N$ is a finite set of odd integers excluding 1, and $\alpha_n, \beta_w, \Tilde{R}_w\in\mathbb{Q}$. In \cite{DHoker:2019blr}, D’Hoker and Green conjectured that the general form of (\ref{eq: RankinSelberg}) involves only numbers of the type \(\ZDG{m} - \ZDG{n}\) with \(\ZDG{n}\) defined by
	\begin{align}
		Z^{\text{DG}}_n=\frac{\zeta_n'}{\zeta_n}-\frac{\zeta_{n-1}'}{\zeta_{n-1}}-\gamma_E \, ,
	\end{align}
	for $n$ a positive integer, and rational numbers that can be expressed through combinations of harmonic sums of integers up to $w+1$. However, this conjecture breaks down at weight 7.
	
	To recover a similar structure, we find that all occurrences of $\gamma_E$ can be absorbed by defining a new variable
	\begin{equation}
		Z_n=\frac{\zeta_{n}'}{\zeta_{n}}-n\gamma_E\, ,
	\end{equation}
	for positive and negative odd integer $n$.  Furthermore, $\gamma_E$ itself can be expressed as the regularized value of $\zpt{1}$
	\begin{equation}
		\gamma_E = \lim_{s\rightarrow 1} \left(\zp{s}+\frac{1}{s-1}\right) \, .
	\end{equation}
	Therefore, up to weight 7, the non-analytic part can be expressed using single-valued MZVs of weight $w-1$, rational numbers, and the logarithmic derivative of the Riemann zeta function. 
	It is tempting to try to absorb the rational numbers into the new variable as well, but doing so requires defining $Z_{-1} - \frac{495541}{200340}$ as our variable at $-1$, which appears unnatural. Thus, we leave the rational numbers independent of $Z_n$. Therefore, we conjecture that (\ref{eq: RankinSelberg}) takes the following general form at all orders
	\begin{align}
		\text{RS}(w) = \mathrm{MZV}^{\mathrm{sv}}_{w-1}\left( \sum_{n\in N} \alpha_n Z_n + \Tilde{R}_w\right)\, .
	\end{align}    
	
	\textit{Transcendentality.—}Finally, we comment on the transcendental behavior of the amplitude. The analytic part is uniformly transcendental as can be seen from (\ref{eq: AnGen}) using conventional assignments of transcendental weights. Furthermore, the $\log$ terms in the non-analytic part of the amplitude multiply single-valued Riemann zeta values of transcendental weight $w-1$, where $w$ is the expansion order. As $\mathcal{T}(\log)=1$, they combine to match the transcendental weights of the analytic part. In \cite{Huang:2024ihm} the coefficients of the logarithmic terms were shown to preserve uniform transcendentality at any genus.
	
	From equations (4.23) and (4.25) of \cite{Green:2008uj}, again using only conventional assignments of transcendental weight, we observe that the supergravity contribution breaks uniform transcendentality by two units.
	
	For the terms without $\log(-s_{ij})$ inside the non-analytic part at order $w$, we observe single-valued MZVs with transcendental weight $w-1$, multiplied with $Z_n$ and rational numbers. To adhere to uniform transcendentality for the $Z_n$ contribution, it is appealing to assign transcendental weight 1 to $Z_n$ based on arguments involving infinite sums such as
	\begin{align}
		\zeta^{\prime}_z = - \sum_{k=1}^\infty \frac{\log(k)}{k^z} \, ,
		\label{eq: derivativeZetasum}
	\end{align}
	which suggests a transcendental weight of $z+1$ of $\zeta^{\prime}_z$, or involving limiting procedures like
	\begin{align}
		\zp{z}=-\lim_{s\rightarrow 1} \left(\frac{\zeta_z\zeta_s}{\zeta_{z+s-1}}-\zeta_s\right) \, ,
		\label{eq: derivativeZetalim}
	\end{align}
	which suggests assigning transcendental weight 1 to $\zp{z}$. This leaves left-over terms $R_w \text{MZV}^{\text{sv}}_{w-1} \sigma_w$ and $\tilde{R}_w \text{MZV}^{\text{sv}}_{w-1} \sigma_w$, where $R_w$ and $\tilde{R}_w \in \mathbb{Q}$, which would break uniform transcendentality by one unit. However, there is no rigorous proof that supports the assignment of transcendental weights to limiting procedures and infinite sums such as (\ref{eq: derivativeZetasum}) and (\ref{eq: derivativeZetalim}). Therefore, we cannot make a conclusive statement on the breaking of uniform transcendentality beyond the supergravity contribution to the non-analytic part of the amplitude.
	
	\textit{Conclusion.—}The results of this letter demonstrate that the four-graviton one-loop closed superstring amplitude, up to 7th order in the low-energy expansion, can be expressed as a $\mathbb{Q}$-linear combination of single-valued multiple zeta values and logarithmic derivatives of Riemann zeta function at odd integers and further suggest a general form of the amplitude. The uniform transcendentality breaks at the supergravity level by two units. {Moreover, the proposed general form of the amplitude suggests a particular transcendental weight assignment to combinations of logarithmic derivatives of the Riemann zeta function.}
	
	Furthermore, in \cite{Huang:2016tag}, it was conjectured that the leading transcendental terms of tree-level string amplitudes are universal. A promising avenue for future research is to extend our approach using modular iterated integrals to compute the analytic and non-analytic parts of the amplitude of different string theories and check whether a similar conjecture is valid at one loop. Additionally, exploring the nature of the numbers appearing in the amplitude and their connections to periods presents an intriguing direction for further study.
	
	\textit{Acknowledgements---}We thank A. Kleinschmidt, O. Schlotterer, and D. Dorigoni for their valuable comments on the draft. We are grateful to G. Bossard, E. D’Hoker, D. Dorigoni, C. Dupont, H. Gangl, M. Green, A. Kleinschmidt, V. Maillot, O. Schlotterer, P. Vanhove, and F. Zerbini for insightful discussions. We also thank the anonymous referees for suggestions on an earlier version. During the last stages of this work, MD’s research was supported by the Munich Institute for Astro-, Particle and BioPhysics (MIAPbP) which is funded by the Deutsche Forschungsgemeinschaft (DFG, German Research Foundation) under Germany’s Excellence Strategy – EXC-2094 – 390783311. MD was also supported by the ERE grant RF$\backslash$ERE$\backslash$221103 associated with the Royal Society University Research Fellowship Grant URF$\backslash$R1$\backslash$221236.
	
%

\end{document}